\documentclass[twocolumn,pra, aps,superscriptaddress,showpacs,groupedaddress]{revtex4-1}
\usepackage{mathptmx}
\usepackage{subfigure}
\usepackage{dcolumn}
\usepackage{amsmath,amssymb}
\usepackage{bm}
\usepackage{color}
\usepackage{latexsym}
\usepackage{epstopdf}
\usepackage{color}
\usepackage[english]{babel}
\usepackage{latexsym}

\usepackage{psfrag,graphicx}
\usepackage{epsf}
\usepackage{subfigure}
\usepackage{amsmath}
\usepackage{amssymb}
\usepackage{amsfonts}
\usepackage{bm}
\usepackage{natbib}
\usepackage{epstopdf}
\usepackage{appendix}
\usepackage{lipsum, babel}

\definecolor{mygrey}{gray}{0.35}
\definecolor{myblue}{rgb}{0.2,0.2,0.8}
\definecolor{myzard}{cmyk}{0,0,0.05,0}
\definecolor{mywhite}{rgb}{1,1,1}
\definecolor{myred}{rgb}{1,0.,0.3}

\usepackage[colorlinks=true,citecolor=myblue,linkcolor=myred]{hyperref}

\def\be{\begin{equation}}
\def\ee{\end{equation}}
\def\ba{\begin{align}}
\def\enda{\end{align}}
\def\bi{\begin{itemize}}
\def\ei{\end{itemize}}

 \def\ee{\mathord{\rm e}}

\def\beq{\begin{equation}}
\def\eeq{\end{equation}}
\def \bml{\begin{multline}}
\def \eml{\end{multline}}
\def \bea{\begin{eqnarray}}
\def \eea{\end{eqnarray}}

\def \ra{{\rightarrow}}

 \newcommand{\ket}[1]{|#1\rangle}
 \newcommand{\bra}[1]{\langle #1|}
 
\newcommand{\bla}[1]{\left(#1\right)}
\newcommand{\blb}[1]{\left[#1\right]}
\newcommand{\blc}[1]{\left|#1\right|}
\newcommand{\bld}[1]{\left\{#1\right\}}

\usepackage{dsfont}

\begin{document}

% The following information is for internal review, please remove them for submission
\widetext

% the following line is for submission, including submission to the arXiv!!
%\hspace{5.2in} \mbox{Fermilab-Pub-04/xxx-E}

\title{Deterministic Quantum Network for Distributed Entanglement and Quantum Computation}
%\title{A single photon deterministically mediates multi-controlled phase gates}
%\title{A multi-controlled phase gate between quantum network nodes is deterministically induced by a single photon}
\author{I. Cohen, and K. M{\o}lmer}
\affiliation{Department of Physics and Astronomy, Aarhus University, Ny Munkegade 120, DK-8000 Aarhus C, Denmark}
%\input author_list.tex       % D0 authors (remove the first 3 lines
                             % of this file prior to submission, they
                             % contain a time stamp for the authorlist)
                             % (includes institutions and visitors)
%\date{\today}

\begin{abstract}
%We propose a simple architecture for a scalable quantum network. The quantum network consists of spins, which are confined in cavities as the quantum nodes, and communicate through a bosonic channel.
We propose a simple interaction protocol to be implemented on a scalable quantum network, in which the quantum nodes consist of qubit systems confined in cavities. The nodes are deterministically coupled by transmission and reflection of a single photon, which is disentangled from the qubits at the end of the coupling operation. This single photon can generate an entangling controlled phase (C-PHASE) gate between any selected number of qubits in the network. Our multi-qubit gate reaches a much higher fidelity compared to schemes concatenating one-qubit and two-qubit gates; thus it forms an efficient basis for universal quantum computing distributed over multiple processor units. %We analyze this network and we show that the gate fidelity is rather increased when more qubits are involved, as requested, e.g., in an efficient Grover search.
In our analysis we consider atomic qubits coupled to optical photons, while the scheme can be readily generalized to other architectures, such as superconducting qubit nodes coupled by microwave photons.
%We propose to generate a multi-controlled phase (CP) gate between N spins, confined in separated cavities. These spins are coupled through a single photon mediator that travels between the cavities, and induces the CP gate without being measured. We analyze this network using the SLH framework, namely, a generalization of the input-output cascade systems, and calculate the gate fidelity, which increases when more spins are involved. As an example we show that our proposal can be utilized as an efficient Grover search platform. In our derivation we consider spins in the optical regime, however, we stress that it can be generalized to many quantum architectures having flying qubits as mediators of the interaction.
\end{abstract}

\pacs{}
\maketitle

%\section{\label{sec:level1}First-level heading}
% sections are not used for PRL papers
%These quantum nodes are connected via quantum channels, enabling the transfer of quantum information and quantum entanglement between the nodes.
%A quantum network locally processes and stores quantum information in quantum nodes that are connected via quantum channels. The quantum information is shared between the nodes by inducing quantum entanglement

{\it Introduction.---}
A quantum network consists of quantum nodes that locally process and store quantum information. The quantum information is then shared between the nodes by linking them via quantum channels \cite{Kimble2008Nature} which mediate the interaction between the nodes \cite{Nickerson2013nc,Nickerson2014prx}. The division of tasks between stationary and flying qubits provides a route to extend quantum computers to operation on large numbers of qubits \cite{Duan2010rmp}, it is at the basis of quantum repeater networks for long distance and multi-user quantum communication \cite{Multi-commu-pra}, and it has applications in metrology \cite{YE_LUKIN_SORENSEN}.  Several theoretical proposals use single photons to couple separated nodes, either in a deterministic \cite{Cirac1997prl}, or a probabilistic (heralded) fashion \cite{DLCZ,Cho2005prl}. Experimental realizations range from atoms \cite{ATOM,GAS01,GAS05,AtomCavity}, trapped ions \cite{ION06,ION07,ION14,Duan2010rmp}, nitrogen-vacancy centers in diamond \cite{NV}, to superconducting qubits \cite{SCQ,  Yale1, Yale2}. So far, experimental studies were restricted to  the coupling of pairs of nodes. While pairwise entanglement and gate schemes are theoretically sufficient to perform general operations on a larger network, this requires the concatenation of operations and drastically limits the experimental feasibility. Here, we propose a new quantum network scheme, in which a single photon is sufficient to mediate a multi-node interaction. Instead of applying a sequence of pairwise operations on quantum nodes consisting of atoms inside optical cavities, our scheme treats the entire network as an interferometer. Subject to the phase shifts incurred under reflection and transmission of a single photon, our scheme
generates a controlled phase (C-PHASE) gate on all qubits involved. The scheme readily lends itself to application on nodes with several atomic qubits, and it should be equally well suited to superconducting circuit architectures with microwave excitation frequencies.

\begin{figure}%[tbp]
\includegraphics[width=8cm]{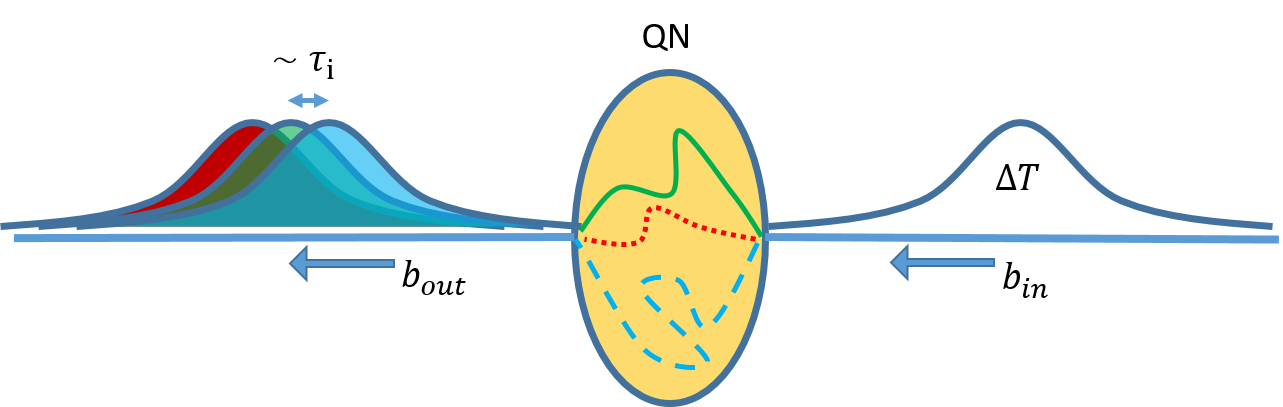}
\caption{A single photon with a long wave packet $\Delta T$  interacts with a quantum network (QN) containing qubits. For different qubit states the photon propagates through different paths and accumulates different phases, giving rise to a multi-qubit phase gate. Depending on the qubit states, the photon wave packet is distorted and delayed (by $\tau_i$), and close to unitary (deterministic) operation on the qubit register requires a high overlap of the different transmitted wave packets. %(b). A network configuration with N qubits (blue circles) located in separated cavities. A C-PHASE gate between the qubits is mediated by a photon entering from the upper right in the figure. (c). Qubit lambda system: the excited state $\ket{e}$ with decay rate $\gamma$ is coupled with strength $g$ to the qubit state $\ket{1}$ by resonant exchange of a photon with the cavity mode, whereas the state qubit $\ket{0}$ is uncoupled. (d). Input-output fields of a one-sided cavity with damping rate $\kappa$, containing a single qubit that is coupled to the cavity mode. (e). Two input and two output fields of a symmetric two-sided cavity $\kappa_L=\kappa_R=\kappa$, containing a single qubit. %(e). Scaling up to a multi-CP gate by placing more two-sided cavities in an optical interferometric set-up. (f.) Scaling to more qubits with registers consisting of $K$ qubits inside each cavity, where a single qubit (blue circle) interacts with the cavity mode and is controlled by the other qubits (black circles).
}
\label{fig1a}
\end{figure}

\begin{figure}%[tbp]
\includegraphics[width=8cm]{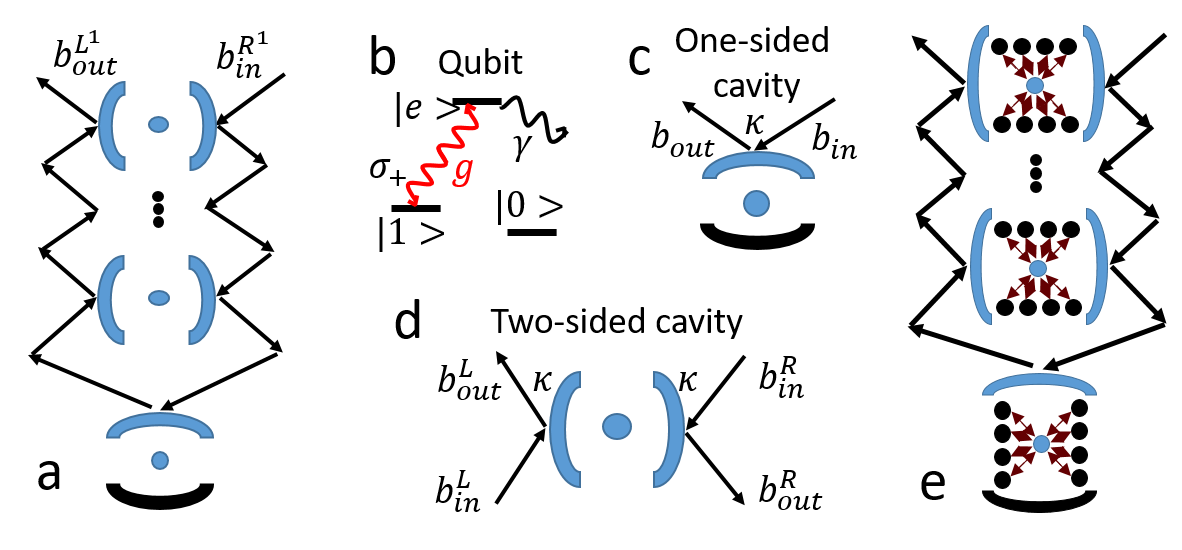}
\caption{ (a). A network configuration with N qubits (blue circles) located in separated cavities. A C-PHASE gate between the qubits is mediated by a photon entering from the upper right in the figure. (b). Qubit lambda system: the excited state $\ket{e}$ with decay rate $\gamma$ is coupled with strength $g$ to the qubit state $\ket{1}$ by resonant exchange of a photon with the cavity mode, whereas the qubit state $\ket{0}$ is uncoupled. (c). Input-output fields of a one-sided cavity with damping rate $\kappa$, containing a single qubit that is coupled to the cavity mode. (d). Two input and two output fields of a symmetric two-sided cavity $\kappa_L=\kappa_R=\kappa$, containing a single qubit. %(e). Scaling up to a multi-CP gate by placing more two-sided cavities in an optical interferometric set-up.
(e.) Scaling to more qubits with registers consisting of $K$ qubits inside each cavity, where a single qubit (blue circle) interacts with the cavity mode and is controlled by the other qubits (black circles).
}
\label{fig1}
\end{figure}

{\it Basic idea.---}
%In our system we consider a single photon having a long wave packet $c\cdot \Delta T$ that propagates through a much smaller quantum network $\sim c\cdot \tau$ (Fig. \ref{fig1a}), with $c$ being the speed of light, and $\Delta T \gg \tau$ are the pulse duration and the temporal delay of the different optical paths inside the quantum network, respectively.
In our system we consider a single photon wave packet that enters and leaves a quantum network through a single input-output channel (Fig. \ref{fig1a}), such that the photon pulse duration  $\Delta T$ is much longer than the temporal delay $\tau_i$ of the $i^{th}$ optical path inside the quantum network. In the $N$ cavity configuration, the quantum network consists of a single one-sided and $N-1$ two-sided cavities containing atomic qubits (Fig. \ref{fig1}a.). We employ interactions, where the cavity mode couples the ground qubit state $\ket{1}$ and excited state $\ket{e}$, and does not couple to the qubit state $\ket{0}$, as shown in Fig. \ref{fig1}b. Hence, the photon is reflected or transmitted from the cavities with phases that depend on the qubit states. In the ideal case, a qubit occupying a $\ket{1}$ state causes the reflection of the photon with a $0$ phase, whereas when a $\ket{0}$ state is occupied, the photon is reflected with a $\pi$ phase from the one-sided cavity,  as demonstrated experimentally in Ref. \cite{CP1}, or transmitted with a $\pi$ phase from the two-sided one. This permits the analysis of the entire network by concatenating the different components in analogy with the analysis of linear classical interferometers, treating each component of the state of the qubits in the product basis $\ket{q_1,q_2, ... ,q_N}$, $q_i=0$ or$1$. The phase factor of this concatenation is a property belonging to the joint quantum state of the photon and the qubits. However, if the photon wave packets depending on the different qubit states sufficiently overlap after the photon leaves the system (see Fig. \ref{fig1a}), the state factorizes and the qubits are disentangled from the photon at the end of the operation. Therefore, we can associate the qubit dependent phase factors with a multi-qubit operation.% alone, where the disentangled photon effectively mediates a many-qubit interaction.

Ideally, in the $N$ cavity configuration (Fig \ref{fig1}a.), when all the $N$ qubits occupy the state $\ket{1}$, the photon reflects from all cavities with no phase shift, and the output single photon wave packet is in phase with the input. Otherwise, the photon is reflected until it encounters the first cavity with a qubit in a $\ket{0}$ state. Here it is transmitted (or reflected if the first $\ket{0}$ qubit state is encountered in the one-sided cavity) with a $\pi$ phase shift. The photon is reflected by its subsequent, second encounter with all the previous cavities since their qubits are still in the $\ket{1}$ states. We thus obtain a change of sign of the one photon wave packet, and the photon mediates the multi-qubit C-PHASE gate between the qubits, where the state $\ket{1}^{\otimes N}$ acquires a $\pi$ phase relative to all other qubit product states.
%The input-output relation of the 50-50 BS is described by the linear relation,
%\beq \begin{pmatrix} b_{out}^{a}  \\  b_{out}^{b}  \end{pmatrix} = \frac{1}{\sqrt{2}} \begin{pmatrix} 1 & i \\ i & 1 \end{pmatrix} \cdot \begin{pmatrix} b_{in}^{a}  \\  b_{in}^{b}  \end{pmatrix}\eeq
%which relates the input and output photon fields of the two sides ($a,b$) of the BS (Fig. \ref{fig1}b.).

{\it Qubit-light interfaces.---} To study the physical case, where the reflection and transmission processes incorporate delays, decays and losses, we shall employ the input-output formalism \cite{InOut1,InOut2}.
The one-sided cavity (Fig. \ref{fig1}c.) with a single qubit with states $\ket{q}$ is illuminated by a single photon, occupying a wave packet mode function, specified by the time dependent amplitude $b_{in}(t)$ arriving at the cavity input mirror.
Using the input-output formalism, detailed in \cite{SI}, we obtain the relation between the input and output wave packets in frequency domain $b_{out}(\omega) =  R_{1q}(\omega)  b_{in}(\omega)$, with the reflection coefficient
 \beq
R_{1,q}(\omega)=\blb{1-\frac{2\kappa\bla{ \gamma -i \omega}}{2\bla{{g_{q}}^2 -\omega^2} +\gamma\bla{\kappa+\kappa' -2 i \omega} - i({\kappa}+\kappa')\omega }}
\label{ref1}
\eeq
Here, $\omega$ denotes the detuning from the cavity resonance, $g_{q}$ is the coupling ($g_0$=0, $g_1=g$) between the qubit and the cavity mode, $\kappa(\kappa')$ is the cavity damping rate by transmission(absorption) losses,
%and the cavity loss rate through the input mirror respectively,
and $\gamma$ is the decay rate of the excited atomic state \cite{definition}. Following the convention in Ref. \cite{CP1,Rempe2018ar}, when $\ket{1}$ is populated, the cavity resonance frequency is split by the strong coupling $g_1= g \gg \gamma, \kappa, \omega$, and the incident photon is reflected at the input mirror with a reflection coefficient $R_{1,1}(\omega)\sim +1$ close to resonance. When the qubit populates the non-coupled state $\ket{0}$ with $g_0=0$, the photon enters the resonant cavity and is reflected with $R_{1,0}(\omega)\sim -1$ close to resonance. %{\color{red} and in the overcoupled cavity case, namely, where the cavity damping rate is dominant over the cavity loss $\kappa \gg \kappa'$ \cite{CP1}}.

Photons far from resonance are reflected with  $R_{1,q}(\omega)\sim +1$ irrespective of the qubit state, emphasizing that our scheme will not work with large frequency bandwidth photon pulses. We note that by writing the solution in the frequency domain we do not assume steady-state driving of the cavity at detuning $\omega$. A Fourier transform back to the time domain yields the build-up and decay dynamics of the cavity excitation amplitude \cite{SI}. In addition to the desired phase shift, the reflection coefficient $R_{1,0}(\omega)$ causes a delay by $\sim 1/\kappa$ of the wave packet, and reduces its amplitude by $\sim \kappa'/\kappa$ due to cavity absorption loss; while $R_{1,1}(\omega)$ reduces the wave packet's amplitude by $\sim \kappa\gamma/g^2$ due to atomic decay, but causes little or no delay. These effects reduce the fidelity of the qubit gate operations and use of our proposal is restricted to long incident photon wave packets and qubit and cavity systems that operate in the high cooperativity ($C \equiv g^2/\kappa\gamma \gg 1$) and overcoupled ($\kappa \gg \kappa'$) regime, which is, indeed the regime explored in experiments \cite{CP1,Rempe2018ar}.

%While the photon is reflected with a phase that depends on the qubit state, we are not supposed to regard this phase factor as a property belonging exclusively to the scattered photon. It is, indeed, a prefactor on the joint state of the qubit and the photon, and the opposite phases $R_{1,q} \simeq \pm 1$ associated with the two qubit states effectively implement a $Z$ phase gate on the atomic qubit, leaving the photon untouched and disentangled from the qubit. The photon may propagate further and interact with other qubits, and thus accumulate further phase shifts that depend on the state of all the qubits. The associated phase factors on the qubit product states, however, merely factorize into a product of single qubit phase gates, and to provide a final phase shift of the photon that enables non-trivial multi-qubit gates, we shall now expand the network with further optical components. Our interferometric set-up, shown in Fig.1.a, with sequential photon interactions with two-sided cavities is inspired by the way that sequences of nearest neighbor two-qubit interactions in a 1D systems mediate gates between remote pairs of qubits. While the photon enters and leaves as a single wave packet, the reader may appreciate that the two-sided cavities endow the photon with a qubit-like path degree of freedom entangled with the atomic qubits inside the set-up.

The two-sided cavity (Fig. \ref{fig1}.d) has input and output ports on both sides of the cavity (index $L,R$). Using the input-output formalism as detailed for the one-photon wave packets in \cite{SI}, we obtain the effective beam splitter relation, which for identical transmission and absorption rates, $\kappa=\kappa_L=\kappa_R$, $\kappa'=\kappa_L'=\kappa_R'$ of the left and right mirror reads
 \beq
%\begin{array}{c}
\begin{pmatrix} b_{out}^{R}(\omega)  \\  b_{out}^{L}(\omega)  \end{pmatrix} = \begin{pmatrix}R_{2,q} & T_{2,q} \\T_{2,q} & R_{2,q} \end{pmatrix}  \begin{pmatrix} b_{in}^{R}(\omega)  \\  b_{in}^{L}(\omega)  \end{pmatrix}\,
%\end{array}
\label{ref1}
\eeq
where the transmission coefficient reads
\beq
T_{2,q}(\omega)={-\frac{\kappa\bla{ \gamma -i \omega}}{\bla{{g_{q}}^2 -\omega^2} +\gamma\bla{\kappa+\kappa'- i \omega} - i({\kappa}+\kappa')\omega }}
\eeq
and the reflection coefficient is $R_{2,q}(\omega)= 1+T_{2,q}(\omega) $. Ideally, the qubit state $\ket{1}$ splits the cavity resonance and prevents entrance of the resonant photon into the cavity, giving rise to $R_{2,1}=1$, and $T_{2,1}=0$; while the inert qubit state $\ket{0}$ corresponds to an empty cavity which transmits a resonant classical field (and hence the wave packet mode function) with a negative amplitude, $T_{2,0}=-1$ and $R_{2,0}=0$ \cite{TWOSIDED}. %{\color{red} Once again, we consider the overcoupled cavity with $\kappa'\ll\kappa$.}
Similar as for the one-sided cavity, delay of the reflected or transmitted wave packet by $\sim 1/\kappa$ should be kept much shorter than the pulse duration and damping should be minimized.
%{\it A network of quantum nodes.---}
%As the photon is reflected $R_{1,q}, R_{2,q}$ or transmitted $T_{2,q}$ with a phase that depends on the qubit states, this phase factor is a property belonging to the joint quantum state of both the photon and the qubits. Ideally, however, the photon is disentangled from the qubits at the end of the operation and we can associate the phase factor with a multi-qubit C-PHASE gate operation.

%In the $N$ cavity configuration (Fig \ref{fig1}a.), when all the $N$ qubits occupy the state $\ket{1}$, the photon reflects from all cavities with no phase shift, and the output single photon wave packet is in phase with the input, $b_{out}^{L_1}=b_{in}^{R_1}$. Otherwise, the photon is reflected until it encounters the first cavity with a qubit in a $\ket{0}$ state. Here it is transmitted (or reflected if the first $\ket{0}$ qubit state is encountered in the one-sided cavity) with a $\pi$ phase shift, and is subsequently reflected by its subsequent, second encounter with all the previous cavities since their qubits are still in the $\ket{1}$ states. We thus obtain a change of sign of the one photon wave packet, $b_{out}^{L_1}=-b_{in}^{R_1}$, and the photon thus mediates the multi-qubit C-PHASE gate between the qubits, where the state $\ket{1}^{\otimes N}$ acquires a $\pi$ phase relative to all other qubit product states.
{\it Interferometer analysis.---}
The complex frequency dependent transmission and reflection coefficients permit calculation of the output field of the set-up in Fig. \ref{fig1}.a, that takes into account the distortion and damping of the photon wave packet. This is done by either summing the amplitude contributions from the multiple photon paths through the system, or by
solving consistently for the field amplitudes by matching the incident and outgoing wave packet amplitudes to the reflection and transmission coefficients at all optical components.
This yields a complex transmission coefficient between the input wave packet at the first input port $R^1$ and the output wave packet at the last output port $L^1$: $b_{out}^{L^1}(\omega)=T\bla{\omega, \bld{g_{q_i}}_1^{N}}  b_{in}^{R^1}(\omega)$,  depending on the $N$ qubit states. For the two-cavity configuration, the transmission coefficient reads:
%we obtain the photon amplitude leaving the upper left output port of the set-up, while accumulating a phase depending on the spin states: $b_{out}^b=R\bla{\omega, g_{q_1},g_{q_2}}  b_{in}^b$ with
\beq
\begin{array}{c}
T\bla{\omega,g_{q_1},g_{q_2}}=\\
\frac{2{g_{q_1}}^2  \blb{ {g_{q_2}}^2 -\bla{\gamma - i \omega}\bla{2\kappa + i \omega}   }
-\bla{\gamma - i \omega}\blb{2 {g_{q_2}}^2  \bla{\kappa + i2 \omega} +  \omega \bla{\gamma - i \omega} \bla{ 2 \omega -i 5 \kappa  } }}
{2{g_{q_1}}^2  \blb{ {g_{q_2}}^2 +\bla{\gamma - i \omega}\bla{2\kappa - i \omega}   }
+\bla{\gamma - i \omega}\blb{2 {g_{q_2}}^2  \bla{\kappa - i \omega} -  \omega \bla{\gamma - i \omega} \bla{ 2 \omega +  i 5 \kappa  } }},
\end{array}
\eeq
where we have assumed the same damping parameters $\kappa$ and $\gamma$ of both cavities and qubit atoms, and (for simplicity of the expression) $\kappa'=0$. One can see that if the qubits populate the qubit product state $\ket{11}$, with couplings ${g_{q_1}}={g_{q_2}}=g \gg \gamma ,\kappa,\omega$, the global state acquires no phase shift whereas in the other three qubit cases, $\ket{10},\ket{01}$ or $\ket{00}$ either $g_{q_1}$, $g_{q_2}$ or both vanish, and the global states acquire a $\pi$ phase change.

%in the optical circuit $\Phi_i=\bla{\omega_0+\omega}\bla{L_i + \delta L_i}/c$, where $\omega_0$ is the bare photon frequency, $c$ is the speed of light, ${L_i + \delta L_i}$ is the $i^{th}$ distance between any two adjacent components, such that $\omega_0 L_i/c =2\pi n$, with integer $n$, and $\delta L_i$ is a fluctuating distance.
%In addition to reflection and transmission losses and phase shifts, we must also take into account the following technical imperfections: 1) losses from the cavity mirrors.

In addition to the delay and loss by cavities, the photon wave packets suffer temporal delay and loss associated with the propagation between the cavities. These are incorporated by introducing a frequency dependent phase factor $e^{\i\omega\tau}$  and an amplitude factor $e^{-\eta}$ for each optical path traversed by the photon \cite{SI}. The photon follows different paths through the network and for the outgoing pulses to be disentangled from the qubits, the propagation delays should all be kept shorter than the duration of the pulse cf. Fig.\ref{fig1a}. %This dephasing mechanism is thus important for broadband (short duration) incident pulses.
Further below, we shall separately discuss qubit gate errors caused by optical phase fluctuations, e.g., due to mirror positioning errors in the different arms of our interferometric set-up.

{\it Fidelity.---}
%We estimate the fidelity of the CP gate as the averaged state fidelity
% \beq F=  \blc{ \frac{1}{d} \sum_{i=1}^d\bra{ph}\otimes\bra{i}CP^\dagger U \ket{i}\otimes\ket{ph}}^2 \eeq
%where $CP$ is the optimal multi-CP gate, $U$ is the photon mediating gate, $\ket{i}$ is the $i^{th}$ state of the qubits spanning a $d$ dimensional Hilbert space, and $\ket{ph}$ is the photonic state, described by a normalized mode function in the frequency domain $\Phi(\omega)$.
The fidelity of the gate is calculated by evaluating the overlap between the various output wave packets $T\bla{\omega,\bld{g_{q_i}}_1^{N}} \Phi^{in}(\omega)$, and a single, normalized, reference function, $\Phi^{ref}(\omega)$. Note that it is convenient to evaluate these overlaps in frequency domain, and while the input mode function $\Phi^{in}(\omega)$ is normalized to unity, the output field may have reduced norm, due to loss of the photon. We conservatively associate photon loss with a complete gate error, and we hence obtain a lower bound for the multi-qubit C-PHASE gate fidelity, averaged over all $d=2^{N}$ register qubit basis states $\ket{q_1,q_2, ... q_{N}}$:

\beq
F_{N}=
\blc{ \begin{array}{c} \frac{1}{2^{N}}\sum_{\bld{q_i}_1^{N}}\int d\omega \, \cdot \quad\quad\quad\quad\quad\quad\quad\quad\quad\quad\quad\quad\\
\quad e^{i\omega \xi} \blc{\Phi^{in}(\omega)}^2  T\bla{\omega,\bld{g_{q_i}}_1^{N}} CP\bla{\bld{q_i}_1^{N}}
\end{array}
 }^2,
\label{FIDELITY}
\eeq
where, for simplicity, we assume the desired reference output photon mode function to be on the form, $\Phi^{ref}(\omega)= \Phi^{in}(\omega) e^{-i\omega \xi}$, and we optimize the expression with respect to the variable $\xi\in\Re$. The ideal multi-qubit C-PHASE unitary operator is described by its action on the qubit product states,  $CP\bla{\bld{1}_1^{N}} = -1$ (all qubits in state $\ket{1}$) or $CP\bla{\bld{q_i}_1^{N}} = 1$ (otherwise).

We have calculated the fidelity for the two, three, and four cavity cases, as function of the following physical parameters: the cooperativity parameter $C=g^2/\kappa\gamma$, a Gaussian incident wave packet $\blc{\Phi(\omega)}^2=\exp \bla{-\omega^2/2\Delta\Omega^2}/\Delta\Omega\sqrt{2\pi}$ with a bandwidth $\Delta\Omega=2\pi/\Delta T$, identical propagation delays $\tau$ between neighboring cavities, identical cavity transmission and absorption loss rates $\kappa_i=\kappa$, $\kappa_i'=\kappa'$ for all $i\in\bld{1,2,...,{N}}$, and identical photon losses between the cavities $\eta$. We provide here the first order expansion of the fidelity in the small parameters $1/C,\Delta\Omega/\kappa,\tau\Delta\Omega, \kappa'/\kappa,\eta \ll1$:  %Fig. \ref{fig2}G present the fidelity and success probability of a heralded gate scheme relying on detection of the transmitted photon. Fig. \ref{fig2}A show the crucial dependence on the cooperativity parameter $C=g^2/\kappa\gamma$, emphasizing the need for strong coupling. The figure also shows approximate analytical expressions,

%In Fig. \ref{fig2}A-G, we present numerical results for the fidelity for the two, three, and four cavity cases, as function of the different physical parameters, assuming a Gaussian incident wave packet $\blc{\Phi(\omega)}^2=\exp \bla{-\omega^2/2\Delta\Omega^2}/\Delta\Omega\sqrt{2\pi}$ with a bandwidth $\Delta\Omega=2\pi/\Delta T$ (Fig. \ref{fig2}B), identical propagation delays $\tau$ between neighboring cavities (Fig. \ref{fig2}C), phase fluctuations $\delta$ (Fig. \ref{fig2}D), identical cavity transmission and absorption loss rates $\kappa_i=\kappa$, $\kappa_i'=\kappa'$ for all $i\in\bld{1,2,...,{N}}$ (Fig. \ref{fig2}E), and identical photon losses between the cavities $\eta$ (Fig. \ref{fig2}F). Fig. \ref{fig2}G present the fidelity and success probability of a heralded gate scheme relying on detection of the transmitted photon. Fig. \ref{fig2}A show the crucial dependence on the cooperativity parameter $C=g^2/\kappa\gamma$, emphasizing the need for strong coupling. The figure also shows approximate analytical expressions,
\beq
\begin{array}{cc}
1 - F_2\approx \\
{2.5/C+ \bla{2.3 /\kappa^2 +1.1 \tau/\kappa +0.95\tau^2}\Delta\Omega^2 +1.8 \kappa'/\kappa +3.7\eta }\\
1-F_3\approx\\
{2.25/C+\bla{1.3 /\kappa^2 +1.5 \tau/\kappa +2.1\tau^2}\Delta\Omega^2 +1.6\kappa'/\kappa +5\eta}\\
1-F_4\approx \\
{1.6/C+\bla{0.76 /\kappa^2 +1.4 \tau/\kappa +2.7\tau^2}\Delta\Omega^2 +1.3\kappa'/\kappa +6.3\eta}.
\end{array}
\label{Fidelity}
\eeq
As expected from our analysis, the optimal value of the adjustable phase shift variable $\xi$ of the reference mode function leading to these expressions represents a suitable median delay with $\sim 1/\kappa$ and $\sim \tau$ contributions \cite{SI}. We recall that with photon pulses of duration longer than $\mu$s, their spatial extent of several hundred meters readily exceeds realistic distances between cavities in laboratories, and the main time delays are caused by the reflection and transmission processes.  %With that restriction, the joint interaction, including arbitrary propagation distances between components is described by the static interferometric analysis.
In \cite{SI} we show by comparison with numerical evaluation of Eq. \ref{FIDELITY} that the analytical expressions provide correct lowest order approximations of the gate fidelities.

There is no principal lower limit to the different terms in Eq.(\ref{Fidelity}). We have arranged the terms in descending order according to typical current experiments. E.g., in \cite{CP1,Rempe2018ar}: $\bld{g,\gamma,\kappa,\kappa'}=\bld{7.9,3,2.3,0.2} \times 2\pi$ MHz  and $\bld{\Delta T,\tau}=\bld{5 \mu s,10 ns}$, leading to a two qubit C-PHASE gate fidelity is $F_2=0.65$.

As the cooperativity parameter $C$ clearly constitutes a main current limitation to the fidelity of our gate and other matter-light interface protocols, it is important to note that this parameter may have more favorable values, e.g., in circuit QED implementations with superconducting qubits and microwave photons. We may also use other mechanisms to control the cavity reflection and transmission of optical photons by atomic qubits, such as recent theoretical proposals, increasing $C$ by applying the collectively enhanced cavity coupling to an ensemble of atoms, which is, in turn, controlled by the Rydberg excitation of a single atom qubit  \cite{Hao2015sr,Das2016pra,Wade2016pra,Felix}. It is thus possible to obtain higher fidelities than the ones pertaining to our single atom example. For example, for $N_{Re}=2500$  Rydberg atoms, the coupling strength is increased by a factor of $\sqrt{N_{Re}}=50$; thus, increasing $\kappa$ by the same factor would result in $F_2>0.99$, perfectly sufficient for the distribution  of entanglement between network nodes \cite{JIANG_LUKIN_SORENSEN}, and even above the threshold for direct implementation of the surface code \cite{surface}. 

It may appear surprising that the fidelity of the multi-qubit C-PHASE gate is higher for increasing number of qubits $N$. The multi-qubit C-PHASE gate fidelity, indeed, increases with $N$ as $F_N\approx1-\bla{4N-3}/2^{N-1}C$ in the $\Delta\Omega \rightarrow 0$ limit. Such favorable scaling is due, in parts, to the fact that only one out of $2^N$ states has a different output than the others. A more crucial observation for applications is the favorable scaling of our multi-qubit gate in comparison with the growing loss of fidelity by sequential application of $O(N)$ two-qubit gates \cite{N_fidelity}.

{\it Dynamical decoupling against phase fluctuations.---}
If the optical paths of the network are stabilized, e.g., with a classical continuous wave beam \cite{phasestabilization}, small phase fluctuations $\delta\ll1$ will reduce the fidelity $F\approx 1- O(\delta^2)$  \cite{SI}. %Simulating the dependence on the phase fluctuations, is done by averaging over normally distributed phase $\phi_i \sim N\bla{0,\delta}$ of each $i^{th}$ optical path of the optical circuit, with a $\delta$ standard deviation \cite{SI}.
Otherwise, we can use the fact that such fluctuations have finite bandwidth and can hence be compensated by a dynamical decoupling approach in a manner, inspired by Refs. \cite{NV,Viola1,Viola2}. This can be done due to the ability to rotate qubits, and to exempt selected cavities from the multi-qubit C-PHASE operation \cite{block_cavity}, in combination with the transmission of two or more single photon wave packets. Let us explain the protocol for the case of two cavities. Assuming $\phi_1$ and $\phi_2$ random phases of the two optical paths between the cavities (see figure in \cite{SI}), our C-PHASE operation yields the unitary operator $U_{CP_2}=\exp\blb{i\pi\ket{1}_1\bra{1}\otimes\ket{1}_2\bra{1} + i\bla{\phi_1+\phi_2}\ket{1}_1\bra{1}}$.
The following sequence yields the C-PHASE gate and refocuses the random phases:
\beq
\Pi_1  \cdot U_{B_2} \cdot \Pi_1 \cdot \Pi_2  \cdot U_{CP_2} \cdot \Pi_2= - e^{i\bla{\phi_1+\phi_2}} e^{i\pi\ket{1}_1\bra{1}\otimes\ket{1}_2\bra{1}},
\eeq
where $\Pi_i$ denotes $\pi$ pulse rotations of the $i^{th}$ qubit, and $U_{B_2}=\exp\blb{i\ket{1}_1\bra{1}\bla{\pi+ \phi_1+\phi_2}}$ results from the transmission of a second photon while detuning the second cavity \cite{block_cavity}. In \cite{SI} we show numerical and analytical results for the accomplishments of the phase cancellation in the $N$ cavity case by refocusing sequences, involving transmission of $N$ photons, while detuning specific cavities.

{\it Universality.---} %Above we have explained how to induce a many body interaction with a single photon.
%Multi-qubit C-PHASE gates together with single qubit operations constitute a universal set of gates for quantum computation \cite{bible}.
Single qubit continuous rotations can be realized using separate driving fields on the qubits, and together with our proposed multi-qubit C-PHASE gates they constitute a universal set of gates for quantum computation \cite{bible}. E.g., a multi-CNOT gate can be performed by operating with Hadamard ($H$) gates on the target qubit, before and after a multi-C-PHASE operation on the target and the selected set of control qubits \cite{block_cavity}. %Hence, any desired entangled state can be prepared, and verified by a Bell test. 
The fidelity of a two-node entangled Bell state prepared this way is exactly $F_2$ calculated for the two-node C-PHASE operation (Eq. \ref{Fidelity}), while neglecting errors of initialization and single qubit $H$ operations \cite{SI}. Note that the fact that the photon wave packet  is much longer than the physical set-up implies that we cannot separate the interaction of the photon with the different qubits in time and, e.g., perform gates on individual qubits between their interactions with the same photon.

{\it Increasing the Hilbert space.---} To expand the number of qubits, we can incorporate
%small registers with $K$ qubits in each cavity  (Fig. \ref{fig2}b.). If they all interact with the cavity in the same way, photon scattering on a single cavity may depend on whether all atoms are in state $\ket{0}$ of if any number of atoms occupies state $\ket{1}$, effectively implementing a local multi-qubit C-PHASE gate (conditioned on the $\ket{0}$ rather than $\ket{1}$ state) \cite{Wade2016pra,Rempe2018ar} by scattering of a single photon.
%A more generally applicable path towards a scalable design with many qubits may implement
local $K$-qubit quantum registers (Fig. \ref{fig1}e.) on which we can perform mutual quantum gates, e.g., by short range Rydberg blockade or state selective contact atomic interactions, motional gates in ion traps, or on-chip gates between superconducting qubits. A single qubit may then be assigned the role of communicating with the other registers via its selective interaction with the cavity field, and our single photon protocol coupling any number out of $N$ such cavities. We note that the local K-qubit registers may also be employed to  correct errors and distill the entanglement provided by the photon scattering \cite{JIANG_LUKIN_SORENSEN}.
Finding the optimum trade-off between the use of fast local gates and relatively slow, but simultaneous, multi-qubit entangling gates presents an interesting challenge for distributed quantum registers \cite{Duan2010rmp}.

Let us conclude the analysis of our proposal by recalling how our multi-qubit C-PHASE on all qubits is, indeed, the only entangling gate needed for an efficient implementation of the Grover search algorithm \cite{Grover}. This is because the Grover algorithm assumes a $\pi$ phase shift on the targeted element $\ket{x^0} \equiv \ket{q_1^0 q_2^0 ... q_N^0}$ relative to all other states, which we obtain by first applying the rotation that takes all qubit $\ket{q_i^0}$ states into $\ket{1_i}$,  then applying the C-PHASE gate, and finally rotating the qubits back. The crucial inversion about the mean in the Grover algorithm is, obtained in a similar manner, by application of Hadamard gates to all qubits before and after the C-PHASE gate \cite{Saffman2011jpb}. %By use of local Rydberg blockade gates in each sub-register, we obtain an efficient implementation of the Grover search on a large, distributed quantum computer.
See  \cite{Saffman2011qip} for the related implementation and application of collective C$_K$-NOT (Toffoli) gates, which together with inter-cavity C$_N$-NOT gates can form the basis of error correction protocols.

{\it Summary.---} We have proposed a new network architecture where a single photon generates a deterministic  multi-qubit C-PHASE gate between qubits embedded in different cavities. We stress that this architecture only relies on the well-established technology of atom-photon interfaces \cite{CP1,Rempe2018ar}. Since a single photon suffices to mediate the multi-qubit C-PHASE gate, the fidelity of our scheme is higher than other schemes involving 1- and 2-node gates alone.
%We have calculated the fidelity of the gate  and analyzed its dependence on the cooperativity parameter, the photon propagation distances and the number of cavities, and
Therefore, our proposed architecture is a promising candidate for distributed quantum computing applications such as implementation of the Grover search algorithm. Our scheme needs single photons as a quantum ressource as a weak coherent state contains both odd and even photon number components, the latter acquiring identical phase factors by the reflection and transmission processes. Fourier limited single photons are becoming readily available, also in architectures that lend themselves to integrate cavities and transmission lines \cite{Zoller2017nat}, and we recall, that although we considered an optical setup with single atoms coupled to optical cavities in our quantitative analysis, our derivation is readily generalized to other quantum platforms that interact with traveling quanta of excitation, e.g, including microwave photons, spin waves and phonons.
%Notably, our quantum network configuration suits not only for a single photon, but for any odd number Fock state, or a photonic minus cat state \cite{SI}, thus softening the requirement to generate a single photon only.

%, including superconducting qubits...

{\it Acknowledgment.---} We thank M. Saffman, F. Motzoi, and A. H. Kiilerich for helpful discussions. The authors acknowledge support from the Villum Foundation and from  the ARL-CDQI program through cooperative agreement W911NF-15-2- 0061. IC acknowledges support from Marie Skłodowska-Curie grant agreement no. 785902.

%\begin{figure}[t]
%\includegraphics[scale=0.8]{figure1}
%\includegraphics[width=0.45\textwidth]{fig1.jpg}
%\caption{ A figure caption. The figure captions are automatically numbered.}
%\end{figure}

%\input acknowledgement.tex   % input acknowledgement

%
% ****** End of file template.aps ******

\newpage
\onecolumngrid
\section{Supplementary Information}

\renewcommand{\thefigure}{S\arabic{figure}}
\setcounter{figure}{0}
\renewcommand{\theequation}{S.\arabic{equation}}
\setcounter{equation}{0}

\subsection*{Photon-qubit interaction}
The basic ingredient in our quantum network is a high fidelity phase gate between a single incoming-outgoing (external) photon and a node composed of a few-level quantum system with qubit degrees of freedom inside a cavity. The gate dynamics can be described using the input-output formalism \cite{InOut1,InOut2} as follows. In the interaction picture eliminating oscillations at the cavity frequency, the non-hermitian Hamiltonian of the photon-matter interface reads
\beq
H=\int d\omega \omega\, a^\dagger(\omega)  a(\omega) + i \int d\omega g_c(\omega) \bla{a^\dagger(\omega)  b - b^\dagger  a(\omega)}+ ig \bla{\ket{e}\bra{1} b - \ket{1}\bra{e} b^\dagger } -i\gamma\ket{e}\bra{e} - i \frac{\kappa'}{2} b^\dagger b
\label{H}
\eeq
where the external photon, detuned by $\omega$ from the cavity resonance, is described by the annihilation and creation operators $a(\omega)$ and $a^\dagger (\omega)$, the cavity mode is described by $b,b^\dagger$, and the qubit state $\ket{1}$ is coupled to the excited state $\ket{e}$ with strength $g$, while the qubit state $\ket{0}$ is not coupled. Instead of presenting the coupling of the qubit and the cavity to baths causing the atomic decay rate $\gamma$ and mirror absorption loss rate $\kappa'$, we incorporate these effects by non-Hermitian evolution terms of the no-jump component of the Monte Carlo wave functions \cite{MonteCarlo}.
Since the Hamiltonian does not cause transitions $\ket{0}\leftrightarrow\ket{1}$, we can write the wave function of the system for both qubit states $\ket{q} \in\bld{\ket{0},\ket{1}}$:
\beq
\ket{\Psi_{0,1}(t)}=\int d\omega \Phi(\omega,t)a^\dagger(\omega)\ket{0_a,0_b,0/1}+C_b(t)b^\dagger \ket{0_a,0_b,0/1} + C_e(t) \ket{0_a,0_b,e}
\label{V}
\eeq
where $0_{a},0_{b}$ correspond to the vacuum of the external field modes and the cavity mode respectively. The wave function is solved with the Schrodinger equation to give the following amplitude differential equations:
\bea
\dot{\Phi}(\omega,t)=- i \omega {\Phi}(\omega,t) + g_c(\omega)  C_b(t)
\label{field} \\
\dot C_b(t)= - \frac{\kappa'}{2} C_b(t) + g_{q} C_e(t) - \int d\omega g_c(\omega)\Phi(\omega,t)  \label{cavity}\\
\dot C_e(t)= -\gamma C_e(t) - g_{q}  C_b(t)
\label{DE}
\eea
with $g_{1}=g$ and $g_{0}=0$.

The differential equation of the external field amplitude (Eq. \ref{field}) is solved by integrating from $t=0$ where the incident photon wave packet has not yet arrived until an arbitrary later time $t$,
\beq
{\Phi}(\omega,t)=e^{- i \omega t} {\Phi}(\omega,0) +g_c(\omega)\int_0^t ds e^{- i \omega \bla{t-s}}  C_b(s).
\label{past}
\eeq
Similarly we may consider the expression in terms of the later time where the photon wave packet has completely left the cavity,
\beq
{\Phi}(\omega,t)=e^{- i \omega \bla{t-t_f}} {\Phi}(\omega,t_f) - g_c(\omega)\int_t^{t_f} ds e^{- i \omega \bla{t-s}}  C_b(s).
\label{future}
\eeq
Integrating these expressions over the frequency domain we obtain
\bea
\frac{1}{\sqrt{2\pi}}\int d\omega {\Phi}(\omega,t)=  b_{in}(t) + \frac{1}{2}\sqrt\kappa C_b(t)
\label{past2}\\
\frac{1}{\sqrt{2\pi}}\int d\omega {\Phi}(\omega,t)=b_{out}(t) - \frac{1}{2}\sqrt\kappa C_b(t),
\label{future}
\eea
in the Markov approximation where $g_c(\omega)\approx\sqrt{\kappa/2\pi}$ is assumed to be a slowly varying function of $\omega$, and $\kappa$ is the cavity damping rate. Here, we introduce the time dependent single photon input and output wave packets, $b_{in}(t)=\int d\omega e^{-i\omega t} {\Phi}(\omega,0) /\sqrt{2\pi}$ and $b_{out}(t)=\int d\omega e^{-i\omega \bla{t-t_f}} {\Phi}(\omega,t_f)/\sqrt{2\pi}$.
The two solutions for the same field amplitude directly provide the input-output relation: $b_{out}(t)=b_{in}(t)+\sqrt\kappa C_b(t)$, and
inserting the external field (Eq. \ref{past2}) into Eq. \ref{cavity}, we get the equation
\beq
\dot C_b(t)=  g_{q} C_e(t)  - \sqrt{\kappa} b_{in}(t) -\frac{\kappa+\kappa'}{2}C_b(t),
\eeq	
showing how the one-photon amplitude of the cavity field builds up as it is driven by the incident photon wave packet.
%Instead of solving the differential equation (Eq. \ref{field}) from the past time to present, namely $ s\in \bla{0,t}$ as was done in Eq. \ref{past}, we integrate on the future time $s\in \bla{t,t_f}$, which is followed by integrating over the frequency domain:
%\beq \frac{1}{\sqrt{2\pi}}\int d\omega {\Phi}(\omega,t)=b_{out}(t) - \frac{1}{2}\sqrt\kappa C_b(t), \label{future} \eeq
%where $b_{out}(t)=\int d\omega e^{-i\omega \bla{t-t_f}} {\Phi}(\omega,t_f)/2\pi$. Together with Eq. \ref{past2} we obtain the input-output relation: $b_{out}(t)=b_{in}(t)+\sqrt\kappa C_b(t)$.

By Fourier transform of the equations for the state amplitudes and the input-output relation to the frequency domain, we obtain:
\bea
-i\omega C_b(\omega)= g_{q} C_e(\omega) - \sqrt{\kappa} b_{in}(\omega) -\frac{\kappa'+\kappa}{2}C_b(\omega)\\
-i\omega C_e(\omega)= -\gamma C_e(\omega) - g_{q}  C_b(\omega)\\
b_{out}(\omega)=b_{in}(\omega)+\sqrt\kappa C_b(\omega)
\label{in-out}
\eea
This set of linear equations is readily solved and yields the reflection coefficient applied in the main text,
\beq
R(\omega,g_{q})=\frac{b_{out}(\omega)}{b_{in}(\omega)}=1-\frac{2\kappa\bla{ \gamma -i \omega}}{2\bla{{g_{q}}^2 -\omega^2} +\gamma\bla{\kappa+\kappa' -2 i \omega} - i\bla{\kappa+\kappa'}\omega }.
\label{ref1}
\eeq
Crucially, to obtain near unit values of $|R(\omega,g_{q})|$ we are obligated to the overcoupled cavity regime, where the cavity transmission loss dominates absorption loss $\kappa \gg\kappa'$, see also \cite{CP1s}. While the solution in the frequency domain contains no time argument, $b_{in}(\omega)$ is the Fourier transform of the time dependent input pulse, and we can return to the time domain by a Fourier transform and observe the build-up and decay dynamics of the cavity excitation amplitude.
%%%%
%Note that the differential equations for the amplitudes (Eq. \ref{DE}) are equal to the Heisenberg equations of the relevant operators, thus the result (Eq. \ref{ref1}) we obtain for the single photon case are correct for the input-output operators: $ {\hat{b}_{out}(\omega)}=R(\omega,g_{q}){\hat{b}_{in}(\omega)}$. Therefore, the state $\ket{1}_{ph}={\hat{b}_{in}(\omega)}\ket{0}_{ph}$ is transmitted to  \beq
%\ket{1}_{ph}\rightarrow R(\omega,g_{q}){\hat{b}^\dagger_{in}(\omega)}\ket{0}_{ph} =  R(\omega,g_{q})\ket{1}_{ph}.
%\eeq
% However, it is also true for for any Fock state of odd number of photons:
%\beq
%\begin{array}{c}
%\ket{2n+1}_{ph}=\frac{1}{\sqrt{(2n+1)!}}\bla{{\hat{b}^\dagger_{in}(\omega)}}^{2n+1}\ket{0}_{ph}\rightarrow \\
%\frac{1}{\sqrt{(2n+1)!}}\bla{ R(\omega,g_{q}) {\hat{b}^\dagger_{in}(\omega)}}^{2n+1}\ket{0}_{ph} = \bla{ R(\omega,g_{q}) }^{2n+1} \ket{2n+1}_{ph} \approx  R(\omega,g_{q})  \ket{2n+1}_{ph}
%\end{array}
%\eeq
%with $R(\omega,g_{q}) \approx \pm1$.
%The solution also holds for the minus cat state $\ket{\alpha}_{ph}-\ket{-\alpha}_{ph}$:
%\beq
%\begin{array}{c}
%\ket{\alpha}_{ph}-\ket{-\alpha}_{ph}=e^{-|\alpha|^2/2} \bla{e^{\alpha\hat{b}^\dagger_{in}(\omega)}-e^{-\alpha\hat{b}^\dagger_{in}(\omega)}    }\ket{0}_{ph}\rightarrow \\
%e^{-|\alpha|^2/2} \bla{e^{ R(\omega,g_{q}) \alpha\hat{b}^\dagger_{in}(\omega)}-e^{- R(\omega,g_{q})\alpha \hat{b}^\dagger_{in}(\omega)}    }\ket{0}_{ph}  \approx R(\omega,g_{q})\bla{ \ket{\alpha}_{ph}-\ket{- \alpha}_{ph}}
%\end{array}
%\eeq

\subsection*{Input-output theory for the two-sided cavity}
In this configuration we introduce two photonic field baths, corresponding to the left and right sides of the cavity: $a_L(\omega) ,a_L^\dagger(\omega) $ and $a_R(\omega) ,a_R^\dagger(\omega) $.
The non-Hermitian Hamiltonian reads
\beq
H=\sum_{j=L,R}\blb{\int d\omega \omega\, a_j^\dagger(\omega)  a_j(\omega) + i \int d\omega g_j(\omega) \bla{a_j^\dagger(\omega)  b - b^\dagger  a_j(\omega)}  }
+ ig \bla{\ket{e}\bra{1} b - \ket{1}\bra{e} b^\dagger } -i\gamma\ket{e}\bra{e}  - i \frac{\kappa_L'+\kappa_R'}{2} b^\dagger b
\eeq
with cavity absorption loss rates associated with both cavity mirrors. Following the arguments from the previous section, the wave function of the two qubit states $\ket{q}\in \bla{\ket{0},\ket{1}}$ is described by
\beq
\ket{\Psi_{0,1}(t)} = \sum_{j=L,R} \blb{\int d\omega \Phi_j(\omega,t)a_j^\dagger(\omega)\ket{0_{a_L},0_{a_R},0_b,0/1}}+C_b(t)b^\dagger \ket{0_{a_L},0_{a_R},0_b,0/1} + C_e(t) \ket{0_{a_L},0_{a_R},0_b,e}
\eeq
where $0_{a_L},0_{a_R}$ correspond to the vacuum of the left and right external field modes respectively.
We obtain a set of differential equations for the amplitudes of the state:
\bea
\dot{\Phi_L}(\omega,t)=-i \omega {\Phi_L}(\omega,t) + g_L(\omega)  C_b(t)
\label{field_L} \\
\dot{\Phi_R}(\omega,t)=-i \omega {\Phi_R}(\omega,t) + g_R(\omega)  C_b(t)
\label{field_R} \\
\dot C_b(t)=-\kappa' C_b(t)+ g_{q} C_e(t) - \int d\omega \blb{ g_L(\omega)\Phi_L(\omega,t)+ g_R(\omega)\Phi_R(\omega,t)}  \label{cavity_j}\\
\dot C_e(t)= -\gamma C_e(t) - g_{q}  C_b(t)
\eea
The differential equations of the external field amplitudes (Eq. \ref{field_L}, \ref{field_R}) are solved by integration through past and future times as before
\bea
{\Phi_j}(\omega,t)=e^{-i\omega \bla{t}} {\Phi_j}(\omega,0) - g_j(\omega)\int_0^{t} ds e^{-i\omega \bla{t-s}}  C_b(s).
\label{past_j}\\
{\Phi_j}(\omega,t)=e^{-i\omega \bla{t-t_f}} {\Phi_j}(\omega,t_f) - g_j(\omega)\int_t^{t_f} ds e^{-i\omega \bla{t-s}}  C_b(s),
\label{future_j}
\eea
for $j=L,R$.
After integrating over the frequency domain we obtain
\bea
\frac{1}{\sqrt{2\pi}}\int d\omega {\Phi_j}(\omega,t)=b^j_{{in}}(t) + \frac{1}{2}\sqrt{\kappa_j} C_b(t)
\label{past2_j}\\
\frac{1}{\sqrt{2\pi}}\int d\omega {\Phi_j}(\omega,t)=b^j_{{out}}(t) - \frac{1}{2}\sqrt{\kappa_j} C_b(t),
\eea
in the Markov approximation. Here, the input and output fields of the $j^{th}$ mode are $b^j_{{in}}(t)=\int d\omega e^{-i\omega t} {\Phi_j}(\omega,0) /\sqrt{2\pi}$ and $b^j_{{out}}(t)=\int d\omega e^{-i\omega \bla{t-t_f}} {\Phi_j}(\omega,t_f)/\sqrt{2\pi}$, providing the input-output relation of the left and right modes: $b^j_{out}(t)=b^j_{in}(t)+\sqrt{\kappa_j} C_b(t)$.
Inserting the external field (Eq. \ref{past2_j}) into Eq. \ref{cavity_j}, we get
\beq
\dot C_b(t)= g_{q} C_e(t) - \sqrt{\kappa_L} b^L_{in}(t)- \sqrt{\kappa_R} b^R_{in}(t) -\frac{\kappa_L+\kappa_R+\kappa_L'+\kappa_R'}{2}C_b(t),
\eeq	
which leads to the linear set of equations in frequency domain:
\bea
-i\omega C_b(\omega)= g_{q} C_e(\omega) - \sqrt{\kappa_L} b^L_{in}(\omega)- \sqrt{\kappa_R} b^R_{in}(\omega) -\frac{\kappa_L+\kappa_R+\kappa_L'+\kappa_R'}{2}C_b(\omega)\\
-i\omega C_e(\omega)= -\gamma C_e(\omega) - g_{q}  C_b(\omega)\\
b^L_{out}(\omega)=b^L_{in}(\omega)+\sqrt{\kappa_L} C_b(\omega)\\
b^R_{out}(\omega)=b^R_{in}(\omega)+\sqrt{\kappa_R} C_b(\omega).
\eea
These equations are solved, and in the symmetric case where $\kappa_L=\kappa_R=\kappa$ and $\kappa_L'=\kappa_R'=\kappa'$, we obtain the reflection and transmission coefficients applied in  the main text:
\beq
b_{out}^{R,L}(\omega)=
 b_{in}^{R,L}(\omega) -    \frac{\kappa\bla{ \gamma -i \omega}}{\bla{{g_{q}}^2 -\omega^2} +\gamma\bla{\kappa+\kappa' - i \omega} - i\bla{\kappa+\kappa'}\omega }     \bla{b_{in}^{R}(\omega)+b_{in}^{L}(\omega)}.
\eeq
Also here we are forced to employ the overcoupled cavity regime where the absorption loss rate is much less than the transmission loss rate $\kappa'\ll\kappa$.
%%%%%

\subsection*{Numerical simulation of the fidelity}
\begin{figure*}
\centering
\includegraphics[width=\textwidth]{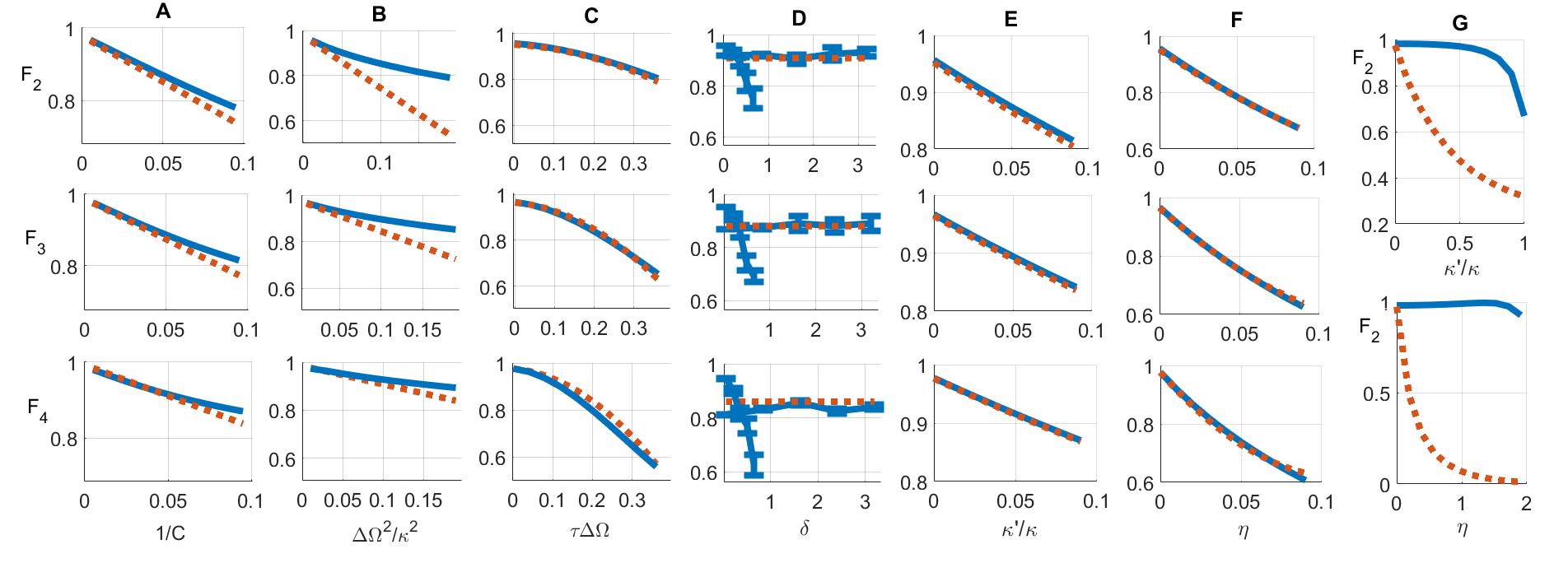}
\caption{  (A) The analytical (red dotted curve) and numerically simulated (blue solid curve) fidelity of the deterministic multi-C-PHASE gates for the two cavity case (upper row), three cavity case (middle row) and four cavity case (bottom row). Unless otherwise specified, the parameters are $C^{-1}=\bla{\Delta\Omega/\kappa}^2=0.01$, $\tau=\delta =\kappa'=\eta=0$ and $\kappa=\bla{2\pi}10$ MHz, and $\gamma=\bla{2\pi}1$ MHz.
The fidelity is shown as a function of (A) the cooperativity $F_N\bla{C^{-1}}$; (B) the photon bandwidth $F_N\bla{\bla{\Delta\Omega/\kappa}^2}$; (C) the propagation delays, $F_N\bla{\tau}$; (D) the phase noise, $F_N\bla{\delta}$ for $\tau=0.01$, without ($\sim 1-O(\delta^2)$ curve) and with dynamical decoupling (constant curve); (E) the mirror losses $F_N\bla{\kappa'/\kappa}$; (F) photon loss in each bosonic channel $F_N\bla{\eta}$. The success probability (red dotted curve) of detecting the photon and thus heralding a two-qubit C-PHASE gate with a high numerically simulated fidelity (blue solid curve) is shown in column G as function of the mirror losses $F_N\bla{\kappa'/\kappa}$ (upper panel) and the photon transmission losses $F_N\bla{\eta}$ (lower panel). Similar results apply to heralded multi-C-PHASE gates for the three and four qubits. %We assume the cavity and qubit damping parameters
}
\label{fig2}
\end{figure*}
In Fig. \ref{fig2}A-G, we present numerical results for the fidelity for the two, three, and four cavity cases, as function of the different physical parameters, assuming a Gaussian incident wave packet $\blc{\Phi(\omega)}^2=\exp \bla{-\omega^2/2\Delta\Omega^2}/\Delta\Omega\sqrt{2\pi}$ with a bandwidth $\Delta\Omega=2\pi/\Delta T$ (Fig. \ref{fig2}B), identical propagation delays $\tau$ between neighboring cavities (Fig. \ref{fig2}C), phase fluctuations $\delta$ (Fig. \ref{fig2}D), identical cavity transmission and absorption loss rates $\kappa_i=\kappa$, $\kappa_i'=\kappa'$ for all $i\in\bld{1,2,...,{N}}$ (Fig. \ref{fig2}E), and identical photon losses between the cavities $\eta$ (Fig. \ref{fig2}F). Fig. \ref{fig2}G present the fidelity and success probability of a heralded gate scheme relying on detection of the transmitted photon. Fig. \ref{fig2}A show the crucial dependence on the cooperativity parameter $C=g^2/\kappa\gamma$, emphasizing the need for strong coupling. The figure also shows approximate analytical expressions (Eq. [6] from main text).

%%%%
\subsection*{Time delay and phase fluctuations}
\begin{figure}[h]
\includegraphics[width=10cm]{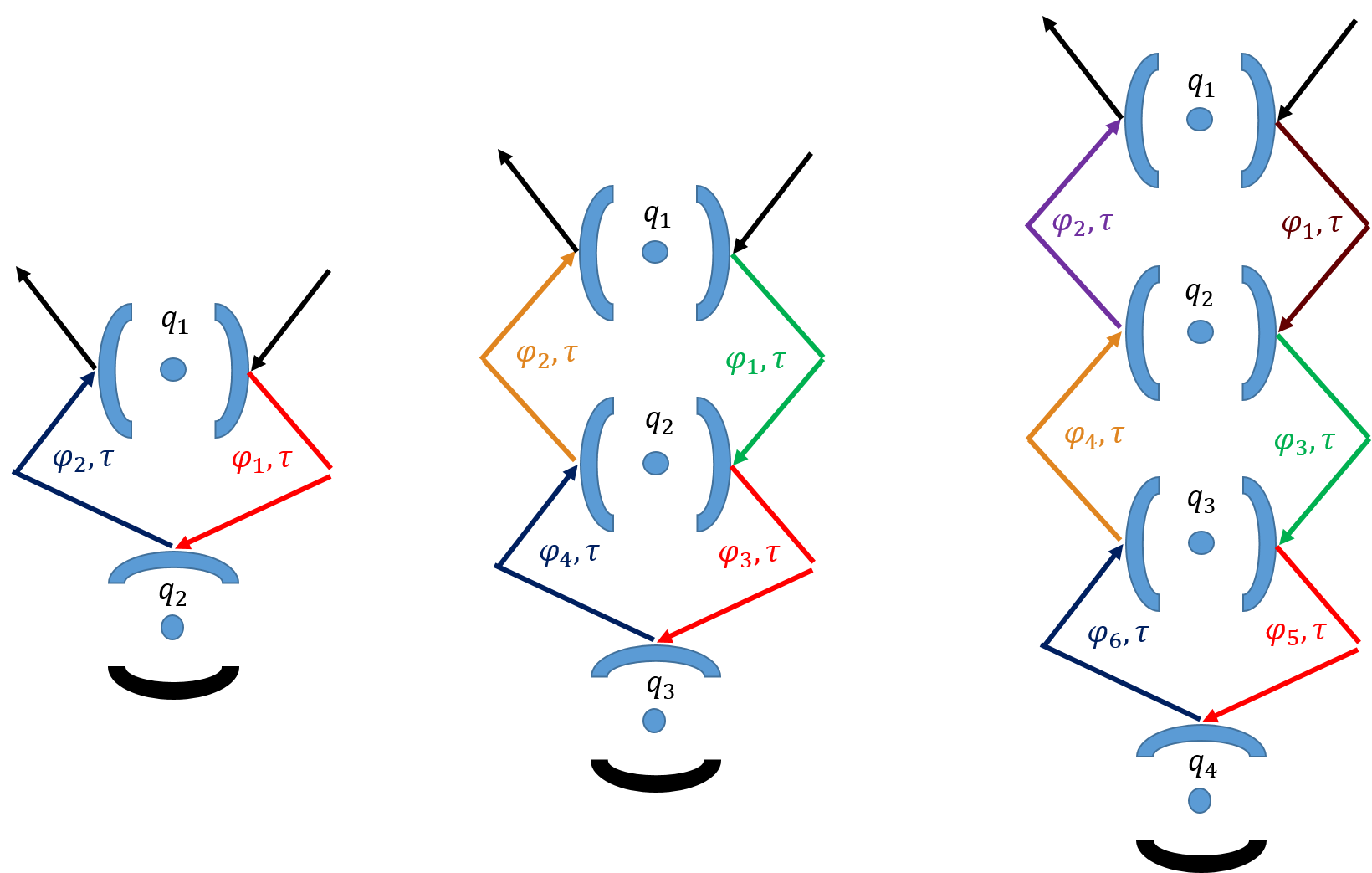}
\caption{Time delays $\tau_i$ and phase fluctuations $\phi_{i}$ associated with the different optical paths.}
\label{fig3}
\end{figure}

We consider equivalent distances between every two adjacent components $\delta L= c \tau$ (Fig. \ref{fig3}). Therefore, the photon is delayed by the corresponding propagation time $\tau$, as represented by a phase factor $\exp\bla{i \tau \omega}$, multiplying the frequency domain wave packet  between every adjacent components. In our analysis, we obtain an underestimate of the fidelity, by calculating the overlap of the transmitted wave packets for all possible qubit settings with a single reference wave, which we only allow to vary from the incident wave by a phase factor $\exp\bla{-i \xi \omega}$. The parameter $\xi$ represents a suitable median delay of the wave packets, and we find its optimal values for $N=$ 2,3 and 4 qubits,
 \beq \begin{array}{cc} \xi_2 = -1.4/\kappa -1.0 \tau\\ \xi_3 = - 1.1/\kappa -1.7 \tau\\ \xi_4 = -0.8/\kappa -2.2 \tau\\ \end{array} \eeq.

In our numerical evaluation of the fidelity we further assume a random phase $\phi_i$ for every $i^{th}$ optical path, distributed normally, $\phi_i \sim N\bla{0,\delta}$, with a standard deviation $\delta$ (Fig. \ref{fig3}).

\subsection*{Generating entanglement}
An entangled state can be generated in the following way. We first initialize the two qubits in the separated cavity nodes in the $\ket{1} $ states, and operate on both with $H$ gates:
\beq
\ket{11} \ra \frac{\ket{00}+\ket{01}+\ket{10}+\ket{11}}{2}
\eeq
We then send a single photon through the network, where we obtain the contribution of the (not perfect) C-PHASE gate:

\beq
\frac{\ket{00}+\ket{01}+\ket{10}+\ket{11}}{2}  \ra  U_{C-PHASE}\frac{\ket{00}+\ket{01}+\ket{10}+\ket{11}}{2} \approx \frac{\ket{00}+\ket{01}+\ket{10}-\ket{11}}{2}
\eeq
Lastly, we apply another $H$ gate on the second (right) qubit where we obtain the entangled state approximately $\ket{\Psi_{ent}}= \blb{\ket{01}-\ket{10}}/\sqrt{2}$. The entangled state fidelity is 
\beq
F_{ent}=\blc{\bra{\Psi_{ent}} H_2 U_{C-PHASE} H_1H_2 \ket{00}}^2
\eeq
Assuming that the main contribution for the loss of fidelity is due to the C-PHASE operation, while neglecting errors of initialization and single-qubit operations such as $H$ gates, the above fidelity reduces to 
\bea
 F_{ent}=\blc{\frac{1}{4}\bla{\bra{00}+\bra{01}+\bra{10}-\bra{11} }U_{C-PHASE}\bla{\ket{00}+\ket{01}+\ket{10}+\ket{11}} }^2=\\
\blc{\frac{1}{4}\bla{\bra{00}U_{C-PHASE}\ket{00}+\bra{01}U_{C-PHASE}\ket{01}+\bra{10}U_{C-PHASE}\ket{10}-\bra{11}U_{C-PHASE}\ket{00} } }^2
\eea
which is exactly the fidelity of the two-node C-PHASE operation $F_2$ from the main paper. 

%We estimate the fidelity of the CP gate as the averaged state fidelity
% \beq F=  \blc{ \frac{1}{d} \sum_{i=1}^d\bra{ph}\otimes\bra{i}CP^\dagger U \ket{i}\otimes\ket{ph}}^2 \eeq  
%where $CP$ is the optimal multi-CP gate, $U$ is the photon mediating gate, $\ket{i}$ is the $i^{th}$ state of the qubits spanning a $d$ dimensional Hilbert space, and $\ket{ph}$ is the photonic state, described by a normalized mode function in the frequency domain $\Phi(\omega)$.

\subsection*{Dynamical decoupling against phase fluctuation}
With the ability to rotate qubits and to exempt selected cavities from the multi-qubit C-PHASE operation, in combination with the transmission of two or more single photon wave packets, we can improve the resilience of our protocol against phase fluctuations similar to the suppression of slowly varying perturbations by dynamical decoupling protocols.

In the case where errors due to finite cooperativity and bandwidth are negligible, the operations needed involve the original, but erroneous,  multi-qubit C-PHASE operation, $U_{CP_N}$, a combination of $\pi$-pulses $\Pi_i$ acting on the $i^{th}$ qubit, and the $U_{B_{\bld{j}}}$ operation caused by the transmission of a photons while blocking one or more $\bld{j}$ cavities.

For the two cavity set-up, these operations read,
\beq
\begin{array}{ccc}
U_{CP}=\exp\blb{i\pi\ket{1}_1\bra{1}\otimes\ket{1}_2\bra{1} + i\bla{\phi_1+\phi_2}\ket{1}_1\bra{1}}\\
U_{B_2}=\exp\blb{i\bla{\pi+ \phi_1+\phi_2}\ket{1}_1\bra{1}}\\
\Pi_i=\exp\blb{i\pi\sigma_{x,i}/2}
\end{array}
\eeq
and the  sequence
\beq
\begin{array}{ccc}
 \bla{ \Pi_1  \cdot U_{B_2} \cdot \Pi_1}  \cdot  \bla{\Pi_2 \cdot U_{CP} \cdot \Pi_2}&=&  e^{i\bla{\pi+ \phi_1+\phi_2}\ket{0}_1\bra{0}} \cdot e^{i\pi\ket{1}_1\bra{1}\otimes\ket{0}_2\bra{0} + i\bla{\phi_1+\phi_2}\ket{1}_1\bra{1}}\\
&=&  e^{i\bla{\phi_1+\phi_2}}  e^{i \pi \ket{0}_1\bra{0}} e^{i\pi\ket{1}_1\bra{1}\otimes\ket{0}_2\bra{0} }\\
%&=&  - e^{-i \pi } \cdot e^{i\bla{\phi_1+\phi_2}} e^{i \pi \ket{0}_1\bra{0}} e^{i\pi\ket{1}_1\bra{1}\otimes\ket{0}_2\bra{0} }\\
%&=&  - e^{-i \pi \blb{\ket{0}_1\bra{0}+\ket{1}_1\bra{1}}} \cdot e^{i\bla{\phi_1+\phi_2}} e^{i \pi \ket{0}_1\bra{0}} e^{i\pi\ket{1}_1\bra{1}\otimes\ket{0}_2\bra{0} }\\
%&=&  - e^{i\bla{\phi_1+\phi_2}}e^{-i \pi \ket{1}_1\bra{1}}   e^{i\pi\ket{1}_1\bra{1}\otimes\ket{0}_2\bra{0} }\\
%&=&  - e^{i\bla{\phi_1+\phi_2}}e^{-i \pi {\ket{1}_1\bra{1}}\otimes\blb{\ket{0}_2\bra{0}+\ket{1}_2\bra{1}}  } e^{i\pi\ket{1}_1\bra{1}\otimes\ket{0}_2\bra{0} }\\
&=&  - e^{i\bla{\phi_1+\phi_2}}e^{-i\pi\ket{1}_1\bra{1}\otimes\ket{1}_2\bra{1}}
\end{array}
\eeq
yields the refocused C-PHASE gate with an unimportant global phase.

In the three cavity case we apply the operations
 \beq
\begin{array}{ccc}
U_{CP}=\exp\blb{i\pi\ket{1}_1\bra{1}\otimes\ket{1}_2\bra{1}\otimes\ket{1}_3\bra{1}+i\bla{\phi_3+\phi_4}\ket{1}_1\bra{1}\otimes\ket{1}_2\bra{1} + i\bla{\phi_1+\phi_2}\ket{1}_1\bra{1}}\\
U_{B_3}=\exp\blb{i\bla{\pi+\phi_3+\phi_4}\ket{1}_1\bra{1}\otimes\ket{1}_2\bra{1} + i\bla{\phi_1+\phi_2}\ket{1}_1\bra{1}}\\
U_{B_{3,1}}=\exp\blb{i\bla{\pi+ \phi_3+\phi_4}\ket{1}_2\bra{1}}\\
\Pi_i=\exp\blb{i\pi\sigma_{x,i}/2},
\end{array}
\eeq
and the dynamical decoupling sequence is
\beq
\begin{array}{ccc}
 \bla{\Pi_2     \cdot U_{B_{3,1}} \cdot  \Pi_2} \cdot \bla{\Pi_1  \cdot U_{B_3} \cdot \Pi_1}  \cdot  \bla{\Pi_3 \cdot U_{CP} \cdot \Pi_3} &=&
e^{i\bla{\pi+ \phi_3+\phi_4}\ket{0}_2\bra{0}}   \cdot e^{i\bla{\pi+\phi_3+\phi_4}\ket{0}_1\bra{0}\otimes\ket{1}_2\bra{1} + i\bla{\phi_1+\phi_2}\ket{0}_1\bra{0}}  \cdot \\
& &\cdot e^{i\pi\ket{1}_1\bra{1}\otimes\ket{1}_2\bra{1}\otimes\ket{0}_3\bra{0}+i\bla{\phi_3+\phi_4}\ket{1}_1\bra{1}\otimes\ket{1}_2\bra{1} + i\bla{\phi_1+\phi_2}\ket{1}_1\bra{1}}\\
&=& e^{ i\bla{\phi_1+\phi_2}\bla{\ket{0}_1\bra{0}+\ket{1}_1\bra{1} } }e^{\bla{\phi_3+\phi_4}\bla{\ket{0}_2\bra{0}+ \ket{0}_1\bra{0}\otimes\ket{1}_2\bra{1} + \ket{1}_1\bra{1}\otimes\ket{1}_2\bra{1} } } \cdot\\
& & e^{i\pi \ket{0}_2\bra{0} } e^{i\pi \ket{0}_1\bra{0}\otimes\ket{1}_2\bra{1}}       e^{i\pi\ket{1}_1\bra{1}\otimes\ket{1}_2\bra{1}\otimes\ket{0}_3\bra{0}}\\
&=& e^{ i\bla{\phi_1+\phi_2}}e^{i\bla{\phi_3+\phi_4}} \cdot e^{i\pi \ket{0}_2\bra{0} } e^{i\pi \ket{0}_1\bra{0}\otimes\ket{1}_2\bra{1}}  e^{i\pi\ket{1}_1\bra{1}\otimes\ket{1}_2\bra{1}\otimes\ket{0}_3\bra{0}}\\
&=& e^{ i\bla{\phi_1+\phi_2}}e^{i\bla{\phi_3+\phi_4}} \cdot \bla{- e^{-i\pi \ket{1}_1\bra{1}\otimes\ket{1}_2\bra{1}}  }   e^{i\pi\ket{1}_1\bra{1}\otimes\ket{1}_2\bra{1}\otimes\ket{0}_3\bra{0}}\\
&=& -e^{ i\bla{\phi_1+\phi_2}}e^{i\bla{\phi_3+\phi_4}} \cdot e^{-i\pi\ket{1}_1\bra{1}\otimes\ket{1}_2\bra{1}\otimes\ket{1}_3\bra{1}}.\\
\end{array}
\eeq

In the $N$ cavity case, the dynamical decoupling sequence consists of $2N$ $\pi$-pulses, and transmission of $N$ photons, giving rise to a refocused multi- C-PHASE, up to a global phase:
\beq
\begin{array}{ccc}
\bla{\Pi_{N-1}     \cdot U_{B_{N,N-2,...,1}} \cdot  \Pi_{N-1}} ...\bla{ \Pi_{i+1} \cdot U_{B_{N,i,...,1}} \cdot  \Pi_{i+1}}...   \cdot\bla{\Pi_2 \cdot U_{B_{N,1}} \cdot \Pi_2} \cdot \bla{\Pi_1  \cdot U_{B_N}\Pi_1} \cdot  \bla{\Pi_N \cdot U_{CP} \cdot \Pi_N} =  \\
=  e^{-i\pi\ket{1}_1\bra{1}^{\otimes N}}\\
\end{array}
\eeq
where the operators are
\beq
 U_{CP}=\exp\blb{\begin{array}{ccc}
i\pi\ket{1}_1\bra{1}^{\otimes N}+\\
i\bla{\phi_{2(N-1)-1}+\phi_{2(N-1)} } \ket{1}_{1}\bra{1}\otimes...\otimes \ket{1}_{N-1}\bra{1}+\\
\vdots\\
i\bla{\phi_{2(N-j)-1}+\phi_{2(N-j)} } \ket{1}_{1}\bra{1}\otimes...\otimes \ket{1}_{N-j}\bra{1}+\\
\vdots\\
i\bla{\phi_{1}+\phi_{2} } \ket{1}_{1}\bra{1}\\
\end{array}  }
\eeq
and
\beq
 U_{B_{N,i,...,1}}=\exp\blb{\begin{array}{ccc}
i\bla{\pi+\phi_{2(N-1)-1}+\phi_{2(N-1)} } \ket{1}_{i+1}\bra{1}\otimes...\otimes \ket{1}_{N-1}\bra{1}+\\
i\bla{\phi_{2(N-2)-1}+\phi_{2(N-2)} } \ket{1}_{i+1}\bra{1}\otimes...\otimes \ket{1}_{N-2}\bra{1}+\\
\vdots\\
i\bla{\phi_{2(N-j)-1}+\phi_{2(N-j)} } \ket{1}_{i+1}\bra{1}\otimes...\otimes \ket{1}_{N-j}\bra{1}+\\
\vdots\\
i\bla{\phi_{2(N-i)+1}+\phi_{2(N-i)+2} } \ket{1}_{i+1}\bra{1}\\
\end{array}  }
\eeq

We subsequently calculate the fidelity of the refocused gates, as function of the bandwidth and cooperativity, with $\kappa'=\eta=0$:
\beq
\begin{array}{cc}
F_2\approx1- \blb{6.5/C+ \bla{2.55/\kappa^2 +0.1 \tau/\kappa +1.95\tau^2}\Delta\Omega^2  }\\
F_3\approx1- \blb{10.25/C+\bla{1.75 /\kappa^2 -0.39 \tau/\kappa +5.4\tau^2}\Delta\Omega^2  }\\
F_4\approx1- \blb{12.625/C+\bla{1.29 /\kappa^2 -0.96 \tau/\kappa +9.0\tau^2}\Delta\Omega^2  },
\end{array}
\eeq
Since the number of photons needed for the multi-C-PHASE gate is increased with $N$, without phase fluctuations the gate fidelity is reduced compared to the single photon gate. See numerical results in Fig. \ref{fig2}D.


\begin{thebibliography}{99}
\bibitem{block_cavity} To choose which qubits are included in the multi-qubit C-PHASE gate we propose to detune the cavities whose qubits shall be excluded from the operation, so they reflect the photon with certainty.
\end{thebibliography}
\end{document}